\documentclass[prl,amsmath,amssymb,floatfix]{revtex4}
\usepackage[latin9]{inputenc}
\usepackage{color}
\usepackage[caption=false]{subfig}
\usepackage{verbatim}
\usepackage{graphicx}
\usepackage{esint}
\usepackage{epstopdf}
\usepackage{tikz-feynman}
\usetikzlibrary{arrows}

\makeatletter
\@ifundefined{textcolor}{}
{%
 \definecolor{BLACK}{gray}{0}
 \definecolor{WHITe}{gray}{1}
 \definecolor{ReD}{rgb}{1,0,0}
 \definecolor{GReeN}{rgb}{0,1,0}
 \definecolor{BLUe}{rgb}{0,0,1}
 \definecolor{CYAN}{cmyk}{1,0,0,0}
 \definecolor{MAGeNTA}{cmyk}{0,1,0,0}
 \definecolor{YeLLOW}{cmyk}{0,0,1,0}
}
\newcommand{\beq}{\begin{equation}}
\newcommand{\eeq}{\end{equation}}

\usepackage{epsfig}
\usepackage{dcolumn}
\usepackage{bm}
\def\(({\left(}
\def\)){\right)}
\def\[[{\left[}
\def\]]{\right]}

\newcommand{\be}{\begin{equation}}
\newcommand{\ee}{\end{equation}}
\newcommand{\bea}{\begin{eqnarray}}
\newcommand{\eea}{\end{eqnarray}}

\makeatother

\begin{document}

\title{ The Eigenstate Thermalization Hypothesis and  Out of Time Order  Correlators}

\author{Laura Foini and Jorge Kurchan}

\address{Laboratoire de Physique Statistique, D\'epartement de physique de l'ENS, Ecole Normale Sup\'erieure, PSL Research University, Universit\'e Paris Diderot, Sorbonne Paris Cit\'e, Sorbonne Universit\'es, UPMC Univ. Paris 06, CNRS, 75005 Paris, France}

\begin{abstract}
The Eigenstate Thermalization Hypothesis (ETH)  implies a form for the matrix elements of local operators between eigenstates of the Hamiltonian, expected to be valid for chaotic systems. Another signal of chaos is a positive Lyapunov exponent, defined
on the basis of Loschmidt echo or out-of-time-order correlators. 
 For this exponent to be positive, correlations between matrix elements unrelated by symmetry, usually neglected,  have to exist.
 The same is true for the peak of the dynamic heterogeneity length $\chi_4$, relevant for systems with slow dynamics. 
 These correlations, as well as those between elements of different operators, are encompassed in a generalized form of ETH.
\end{abstract}

\maketitle

\section{Introduction}

In  recent years there has been a renewal of interest in chaotic
systems within a quantum mechanical description.
A characterization of wavefunctions is the Eigenstate Thermalization Hypothesis (ETH), 
an assumption that leads to a form for
the matrix elements of local operators in the energy eigenbasis and 
can be viewed as an extension of ideas borrowed from random matrix theory.
On the other hand,  the Lyapunov exponent, a fingerprint of chaoticity in classical systems describing the
sensitivity of the system's dynamics to small perturbation \cite{pasta}, has been
considered also for quantum systems based on  the definition of suitable four point correlation functions.
Both the applicability of the ETH to chaotic systems and  the emergence
of a Lyapunov exponent in their dynamics are  subjects of active current research.
The simultaneous consideration  of these two aspects for non-integrable systems raises the
question of how high order correlation functions are described within the ETH ansatz, the subject of this paper.

 In the following we consider a hamiltonian $H$ which we assume is not integrable, and its eigenvalues  $e_i$. 
We assume that there is a parameter (e.g. $\hbar$ or the number of degrees of freedom $N$) that controls the level spacing
$\delta e$. Because timescales  and energy are inversely related $t \delta e \sim \hbar$, this also defines the Heisenberg time $2 \pi \hbar / \delta e$, the longest in the system's evolution (`a theoreticians's time' \cite{srednicki_99}).
Between the smallest microscopic time and the Heisenberg time are  physical times  where some correlations are non-zero \cite{foot}. 
Because the level density is $\sim \hbar^N$ in the  semiclassical limit,  both small $\hbar$ and large $N$  may be  cases where the separation of correlation
and Heisenberg times is large, and we are thus allowed to consider  an energy scale $\Delta$ that contains many levels, but is much smaller than
the energy corresponding to the inverse of all interesting physical times.

The  Eigenstate Thermalization Hypothesis, as applied to expectation values \cite{deutsch_91, srednicki_94,srednicki_99,kafri}
 is usually stated as follows:  for the matrix elements of a smooth observable $A$ { in the energy eigenbasis}, putting $e_+ \equiv (e_i+e_j)/2$:
\begin{equation}
A_{ij} = a(e) \delta_{ij} + R_{ij} F(e_+, e_i-e_j) e^{-S(e_+)/2}
\label{ETH}
\end{equation}
where  both $ a(e)$ and $F(e_+,\omega)$ are assumed to be smooth functions of their arguments.  For macroscopic systems, as we shall mostly consider here, 
we may assume that $a(e)$ is a smooth function of the energy {\em density}. The origin of the factor $e^{-S(e_+)/2}$  comes from the spreading of weights over many levels, and is the only
one compatible with $F(e_+,\omega)$ of order one, as we shall see. The $|R_{ij}|$ have expectation of order unity. 
Here  we may consider $A$ that are  real or complex hermitian. We have
\begin{eqnarray} 
R_{ij}&=&R_{ji} \;\; {\mbox{real}} \;\;\; ; \;\;\; \;\; \;\; {\mbox{for $ A$ real}} \nonumber\\
R_{ij}&=&R^*_{ji} \;\; {\mbox{complex with random phase} } \;\;\; ; \;\;\; \;\;  \;\; {\mbox{for $A$ complex}}
\label{realcomp}
\end{eqnarray}

Now, the off-diagonal elements $R_{ij}$  
are usually considered as independent  random variables and therefore their product, apart from the obvious symmetries above, average to zero. Does this assumption together with (\ref{ETH}) 
contain everything that is necessary to calculate $n$-point functions? 
Clearly not, as this would ultimately  mean that all higher correlation functions are obtainable as model-independent functionals of the
one and two-point functions, i.e. all models would be Gaussian. 
 {\em In particular the  correlations that determine the Lyapunov exponent \cite{larkin,mss,kitaev,pasta,loschmidt1,loschmidt2,PZ}, as well as the dynamic heterogeneities $\chi_4$, are not fully included 
 -- and the most important physical part is indeed absent -- in (\ref{ETH})}.

Let us now be more clear on what ensemble we have in mind when we say `random'. We may consider
{\em i)}  energies averaged over  windows of extension in energy $\Delta$ as above, {\em ii)} a perturbation $H \rightarrow H +  \Delta  H_r$, where
$H_r$ is a local random Hamiltonian and the coupling is strong enough to couple nearby levels but sufficiently weak as to not affect correlation functions at physical times (see Deutsch \cite{deutsch_91}), {\em iii)} a set of transformations $U$ over physical observables described below (`typicality') (see  \cite{neumann,goldstein,reimann2,reimann1}).
Denoting with an  overbar  averages over any of these  ensembles, we may write:
\begin{equation}
\overline{A_{i_1 i_1} } =  \; {\bf F}_{e_{ +}}^{(1)}  \;\;\;\; ; \;\;\;\;  \overline{A_{i_1 i_2} A_{i_2 i_1}  } =e^{-S(e_+)}  \; {\bf F}^{(2)}_{e_+} (e_{i_1}-e_{i_2}) \ \ \ \text{for $i_1\neq i_2$}\ .
\label{etheth1}
\end{equation}
where we have defined $ {\bf F}^{(2)}_{e_+} (e_{i_1}-e_{i_2})=|F(e_+, e_{ i_1}-e_{ i_2})|^2$.

In the following sections we introduce and  justify an  ansatz for higher correlations, and we show how it applies to the calculation of the out of time order correlators (OTOC). We also give  numerical support to the contention that the correlations between matrix elements are not negligible. 

\vspace{.2cm}

{\bf A convenient time-axis}

\vspace{.2cm}

Here we show how working on the shifted time-axis (and correspondingly, a split Gibbs measure in the trace) is the natural strategy \cite{mss,mukund}.
A way to introduce this construction is to consider  two-point functions. 
The correlations and response functions are defined as  (see e.g. \cite{kamenev}):
\begin{eqnarray}
C(t,t') &=& \frac 12 {\mbox{Tr}}\left[  [A(t),A(t')]_+ e^{-\beta H} \right] / Z
\nonumber \\
R(t,t')  &=& \frac  i \hbar  {\mbox{Tr}}\left[  [A(t),A(t')]_-  e^{-\beta H} \right]  \Theta(t-t') / Z
\end{eqnarray}
where we have made explicit $\hbar$.
For $t_1>t_2$:
\begin{equation}
\left\{C-\frac{i\hbar}{2} R\right\}(t_1-t_2)= {\mbox{Tr}}\left[ A(t_1)A(t_2) e^{-\beta H}\right] / Z
\end{equation} 
It is easy to derive the so-called `KMS condition' \cite{KMS}, the quantum version of fluctuation-dissipation relations,
which in the time-domain may be expressed as a symmetry property:
\begin{equation}
\left[\left\{C{-}\frac{i\hbar}{2} R\right\}\left(\tau-\frac{i\beta\hbar}{2}\right)\right]^*=\left\{C{ -}\frac{i\hbar}{2} R\right\}\left(\tau-\frac{i\beta\hbar}{2}\right)
\label{kmss}
\end{equation} 
for all $\tau$.
This strongly suggests that life becomes simpler if one displaces the $t$-axis  by $\frac{i\beta\hbar}{2}$ and defines:
\begin{equation}
O_\beta(t,t') = {\mbox{Tr}}\left[  A(t) e^{-\frac \beta 2 H}  A(t') e^{-\frac \beta 2 H} \right] / Z
\label{tpf}
\end{equation}
The KMS condition (\ref{kmss}) is now the statement that $O_\beta(t,t')$ is real, which is immediate from Hermiticity and the cyclic property of the trace:
\begin{equation}
O_\beta(t,t') = [O_\beta(t,t')]^{\ast} = [O_\beta(t',t)]^{\ast}
\label{kms2}
\end{equation}
We shall see the generalization of this below.

 \vspace{.2cm}

{\bf From matrix elements to correlations}

\vspace{.2cm}

Let us calculate  {\em for times small compared to the Heisenberg time} 
the correlator Eq. (\ref{tpf}), using the ansatz (\ref{ETH}). We shall do this quickly here, and will justify 
the averaging involved in much more detail in the sections below.
\begin{equation}
O_\beta(t,t') = \sum_i a^2(e_i) e^{-\beta e_i}/Z   + \sum_{ij} e^{-\beta(e_i+e_j)/2-S(e_+)} |R_{ij}|^2 |F(e_+, e_i-e_j) |^2  e^{it(e_i-e_j)}/Z
\label{eth1}
\end{equation} 

We may hence change variables $e_+=(e_i+e_j)/2$ and $\omega=e_i-e_j$.
 replacing sums by integrals (see e.g., Ref \cite{srednicki_99,kafri})
\begin{equation} 
\sum_i \; \bullet \;\;\; \rightarrow \int de \; e^{S(e)} \; \bullet
\label{bullet}
\end{equation}
and performing saddle point integration for $e_+$  (we are assuming a thermodynamic system), we obtain
  \begin{eqnarray}
O_\beta(t,t') &=& \int de \; a^2(e) e^{-\beta e + S(e)}/Z   + \int d e_+ d \omega \;e^{-\beta e_++S(e_+)} |F(e_+, \omega) |^2  e^{it\omega}/Z \nonumber \\
&=& \left[ {\bf F}^{(1)}(e_\beta) \right]^2+ \int d \omega {\bf F}^{(2)}_{e_\beta}( \omega)   e^{it\omega}
\label{correl}
\end{eqnarray} 
where $e_\beta$ is the canonical energy at temperature $\beta$.

We first observe that the diagonal term must be of order one if $\lim_{t \rightarrow \infty} O(t) =a$ is finite, while the scale $e^{-S(e_+)}$ is necessary
so that the result is of order one, as it should. 
In conclusion, we find that the function $|F|^2(\omega)= {\bf F}^{(2)}_{e_\beta}( \omega) $ is the Fourier transform of the correlation (\ref{tpf}) {\em which is defined along the shifted time axis}. The usual correlation and response
are obtained by shifting back the time-axis, and in so doing collecting a factor $\cosh (\hbar \beta \omega /2)$ and $\sinh (\hbar \beta \omega /2)$, respectively \cite{kafri}.
Indeed, splitting the Gibbs measure gives a much simpler relation between time correlators and correlations between matrix elements, and this will be even more so in the sequel.

\section{Matrix elements and many-time correlation functions.}\label{Sec_matrix_elements}

In Section \ref{Sec_tipicality} we shall show, using typicality considerations, that expectations of products of matrix elements
in which  indices are not repeated  vanish. In this section we shall
also { argue } that if an index is repeated more than twice, then the expectation breaks into a product of expectations.
All in all, { for the purposes of computing physical correlation functions,} 
we shall only need to discuss the following generalization of (\ref{etheth1}) for $i_1,...,i_n$ {\em different}:
\begin{equation}
\overline{A_{i_1 i_2} A_{i_2 i_3} ... A_{i_n,i_1} } = e^{-(n-1)S(e_+)} \; {\bf F}^{(n)}_{e_+} (e_{i_1}-e_{i_2}, ..., e_{i_{n-1}}-e_{i_n}) 
\label{etheth}
\end{equation}
where $e_+=(e_{i_1}+e_{i_2}+ ...  +e_{i_n})/n$ and $ {\bf F}^{(n)}_{e_+} (e_{i_1}-e_{i_2}, ..., e_{i_{n-1}}-e_{i_n})$ is a smooth function of the energy differences and $a(e_+)$ a 
a smooth function of the average energy
{\em density} $e_+/N$.
The factor:
\begin{equation}
{\cal{N}}^{(n)} \equiv e^{-(n-1)S(e_+)}
\label{norma}
\end{equation}
guarantees that ${\bf F}^{(n)}$ is of order one. Averages are as above. 
Expression (\ref{etheth}) contains (\ref{etheth1}) as a particular case for  $n=1,2$. A similar expression may be expected to hold if the matrices are different (e.g. $A_{ij}B_{ji}$), with an appropriate
function ${\bf F}^{(n)}$ dependent on the operators involved.

 Note that the usual ETH assumes an  absolute value of the $A$'s of the order $e^{-S(e_+)/2}$, while the part contributing to higher correlations is much smaller: $\sim e^{- \frac{n-1}{n} S(e_+)}$ {\em per factor}.
  These correlations between $n>2$ matrix elements  are indeed smaller, 
but enter all expressions for $n$-point time-correlations through sums with many more terms, so that their contribution 
to physical correlation functions
 is of the same order -- { just as what happens}  between diagonal and off-diagonal
terms in (\ref{etheth1}). { Indeed, as we shall see explicitly in the next subsections, a sum over $n$ indices 
of the average (\ref{etheth}) picks up a factor $\sim e^{nS}$ from the density of levels, while we may estimate the sum of the (larger) fluctuating parts as being only a factor $\sim e^{nS/2}$, due to random signs.}

We have restricted ourselves to different indices. 
In fact, when there is a repetition one can see that what dominates are products of lower order
expectations { under some assumptions on the fluctuating parts}.. Consider, for example the case in which $i_r=i_1$, and all other indices are different:
\begin{eqnarray}
& & \underbrace{A_{i_1, i_2} A_{i_2 i_3} ... A_{i_{r-1},i_1}}  \underbrace{A_{i_1,i_{r+1}} ... A_{i_n,i_1}} \equiv {\cal C}
{\cal D}=\overline{A_{i_1 i_2} A_{i_2 i_3} ... A_{i_{r-1}i_1}} \;\;\;  \overline{A_{i_1i_{r+1}} ... A_{i_n,i_1}} +\tilde {\cal C}\tilde {\cal D}
\label{cuatro}
\end{eqnarray}
where we have splitted into  average and fluctuating parts $ {\cal C}= \overline {\cal C}+ \tilde {\cal C}$ and  $ {\cal D}= \overline {\cal D}+   \tilde {\cal D}$.
We assume that the  average of fluctuating parts $\overline{\tilde {\cal C}\tilde {\cal D} 
}$ scales at most like (\ref{etheth}) (i.e. as the result without index repetitions).
We wish to compare this  with the product of averages, which  contain $(r-1)$ and $(n-r+1)$ terms, respectively.
Put $e_a= \frac{e_{i_1}+... + e_{i_{r-1}}}{(r-1)}$ and $e_b= \frac{e_{i_r}+... + e_{i_{n}}}{(n-r+1)}$, so that:
$n e_+=(r-1) e_a+(n-r+1)e_b$.
The size of the estimate (\ref{etheth}) is the exponential of
\begin{equation}\label{entropy}
 - (n-1) S (e_+) = -(n-1) S\left[\frac{(r-1) e_a+(n-r+1)e_b}{n}\right] \le -(n-1) \left[ \frac{(r-1)}{n} S(e_a) + \frac{(n-r+1)}{n} S(e_b) \right]
 \end{equation}
where in the inequality we have used the convexity of $S$ (itself a consequence of the positivity of specific heat).
This is to be compared with the estimate for the logarithm of the product:
\begin{equation}
 \ln ({\cal{N}}^{(r-1)} {\cal{N}}^{(n-r+1)})=  -(r-2)  S(e_a) - (n-r)  S(e_b) 
 \end{equation}
Using the fact that $n-r+1$ and $r-1$ are nonnegative, we check that the product  form
always dominates
\begin{eqnarray}
& & \overline{{A_{i_1, i_2} A_{i_2 i_3} ... A_{i_{r-1},i_1}} {A_{i_1,i_{r+1}} ... A_{i_n,i_1}}} \sim \overline{A_{i_1 i_2} A_{i_2 i_3} ... A_{i_{r-1}i_1}} \;\;\;  \overline{A_{i_1i_{r+1}} ... A_{i_n,i_1}} 
\label{cuatrocuatro}
\end{eqnarray}
In short, $A_{i_1 i_2} A_{i_2 i_3} ... A_{i_{r-1}i_1}$ and  $A_{i_1 i_{r+1}} ... A_{i_n,i_1}$ are stochastic variables whose
 expectations  factorize up to exponential accuracy { if their covariances scale as (\ref{etheth}) }. Their variances are always large, but contribute subdominantly to the sums over indices involved in the calculation of correlation functions.

\begin{figure}[h]
\centering
\begin{tikzpicture}[
roundnode/.style={circle, fill=black!100,minimum size=1mm,scale=0.5},
]
\node[roundnode](i0){};
\node[roundnode](i1)[above right=0.66cm of i0] {};
\node[roundnode](i2)[below right=0.66cm of i1] {};
\node[roundnode](i3)[below right=0.66cm of i0] {};
\draw[->] (i0) arc (180:135:0.6cm);
\draw (i1) arc (90:135:0.6cm);
\draw (i0) arc (180:225:0.6cm);
\draw[->] (i3) arc (270:225:0.6cm);
\draw[->] (i1) arc (90:45:0.6cm);
\draw (i2) arc (0:45:0.6cm);
\draw[->] (i2) arc (0:-45:0.6cm);
\draw (i3) arc (270:315:0.6cm);
\end{tikzpicture}
(a)

\vspace{0.2cm}
\begin{tikzpicture}[
roundnode/.style={circle, fill=black!100,minimum size=1mm,scale=0.5},
]
\node[roundnode](i0){};
\node[roundnode](i1)[below right=0.66cm of i0] {};
\node[roundnode](i2)[above right=0.66cm of i1] {};
\draw[->] (i0) arc (180:90:0.6cm);
\draw (i2) arc (0:90:0.6cm);
\draw[->] (i2) arc (0:-45:0.6cm);
\draw (i1) arc (270:315:0.6cm);
\draw[->] (i1) arc (270:225:0.6cm);
\draw (i0) arc (180:225:0.6cm);
\draw (i2) arc (180:-180:0.3cm);
\end{tikzpicture}
(b)

\vspace{0.2cm}
\begin{tikzpicture}[
roundnode/.style={circle, fill=black!100,minimum size=1mm,scale=0.5},
]
\node[roundnode](i0){};
\node[roundnode](i1)[right=0.8cm of i0] {};
\node[roundnode](i2)[right=0.8cm of i1] {};
\draw[->] (i0) arc (180:90:0.5cm);
\draw (i1) arc (0:90:0.5cm);
\draw (i0) arc (180:270:0.5cm);
\draw[->] (i1) arc (360:270:0.5cm);
\draw[->] (i1) arc (180:90:0.5cm);
\draw (i2) arc (0:90:0.5cm);
\draw (i1) arc (180:270:0.5cm);
\draw[->] (i2) arc (360:270:0.5cm);
\end{tikzpicture}
(c)

\vspace{0.2cm}
\begin{tikzpicture}[
roundnode/.style={circle, fill=black!100,minimum size=1mm,scale=0.5},
]
\node[roundnode](i0){};
\node[roundnode](i1)[right=0.8cm of i0] {};
\draw[->] (i0) arc (180:90:0.5cm);
\draw (i1) arc (0:90:0.5cm);
\draw (i0) arc (180:270:0.5cm);
\draw[->] (i1) arc (360:270:0.5cm);
\draw (i0) arc (0:360:0.2cm);
\draw (i0) arc (0:360:0.3cm);
\end{tikzpicture}
(d)

\vspace{0.2cm}
\begin{tikzpicture}[
roundnode/.style={circle, fill=black!100,minimum size=1mm,scale=0.5},
]
\node[roundnode](i0){};
\node[roundnode](i1)[right=0.8cm of i0] {};
\draw[->] (i0) arc (180:90:0.5cm);
\draw (i1) arc (0:90:0.5cm);
\draw (i0) arc (180:270:0.5cm);
\draw[->] (i1) arc (360:270:0.5cm);
\draw (i1) arc (180:-180:0.3cm);
\draw (i0) arc (0:360:0.3cm);
\end{tikzpicture}
(e)

\vspace{0.2cm}
\begin{tikzpicture}[
roundnode/.style={circle, fill=black!100,minimum size=1mm,scale=0.5},
]
\node[roundnode](i0){};
\node[roundnode](i1)[right=0.8cm of i0] {};
\draw (i0)[->] arc (180:90:0.5cm);
\draw (i1) arc (0:90:0.5cm);
\draw[->] (i1) arc (0:-90:0.5cm);
\draw (i0) arc (180:270:0.5cm);

\draw(i0) arc (135:90:0.7cm);
\draw[->]  (i1) arc (45:90:0.7cm);
\draw[->] (i0) arc (225:270:0.7cm);
\draw (i1) arc (315:270:0.7cm);
\end{tikzpicture}
(f)

\vspace{0.2cm}
\begin{tikzpicture}[
roundnode/.style={circle, fill=black!100,minimum size=1mm,scale=0.5},
]
\node[roundnode](i0){};
\draw (i0) arc (360:180:0.5cm);
\draw (i0) arc (0:180:0.5cm);
\draw (i0) arc (180:360:0.5cm);
\draw (i0) arc (180:0:0.5cm);
\draw (i0) arc (360:180:0.3cm);
\draw (i0) arc (0:180:0.3cm);
\draw (i0) arc (180:360:0.3cm);
\draw (i0) arc (180:0:0.3cm);
\end{tikzpicture}
(g)

\caption{Diagrams corresponding to four-point functions. All the diagrams are cacti apart the (f).}\label{Fig_four_point}
\end{figure}
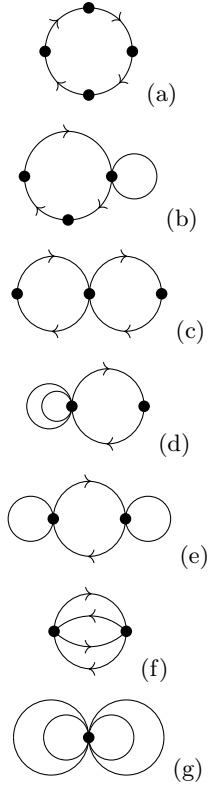

 Proceeding this way, one may  use this argument to generalize (\ref{cuatrocuatro}). In particular,
 the fact that the ansatz (\ref{etheth}) is all we need to calculate $n$ point correlations may be argued as follows.
Consider a product $A_{i_1 j_2} \dots A_{i_n j_n}$. We may represent it as a diagram by using as vertices
all the {\em different} indices amongst the set $\{ i_1, ..., i_n, j_1,...,j_n\}$. Each factor $A_{ij}$ is then represented by an arrow { (for complex $A_{ij}$, real $A_{ij}$ are represented by a simple edge)}
going from $i$ to $j$. 
We consider:  {\em i)} the simplest non-zero expectation, the one involved in ansatz (\ref{etheth}) corresponds to a diagram
with all distinct vertices lying on a single loop.  {\em ii)} Next, we consider { a tree of loops, joined to one  another at single vertices}: a `cactus' diagram (a two-leaf cactus is (\ref{cuatrocuatro})).
The entropic argument  (\ref{entropy})  may be generalized (see Appendix ~\ref{Appendix_entropy}) to argue that
the expectation of such a product is dominated by a product of its constituent loops, each obtained from (\ref{etheth}), 
{under the assumption that the covariaces scale as (\ref{etheth})}. This is the generalization of (\ref{cuatrocuatro}).
{\em iii)} Finally, we have products that are non cacti, for example  $A_{ij}A_{jk}A_{ki}A_{ik}A_{kj}A_{ji}$. These products
may be decomposed in loops in more than one way  (for example $(A_{ij}A_{jk}A_{ki})(A_{ik}A_{kj}A_{ji})=(A_{ij}A_{ji})(A_{ki}A_{ki})(A_{jk}A_{kj})
$. For these we make the assumption that their order
in $e^{-S}$ is the same as the one of the decomposition in loops that dominates. These expectations may be large in order, but they have less free indices that
the corresponding cactus, and hence pick up less factors $e^{S}$ and their contribution to the n-point correlation functions is subdominant.   In Fig.~\ref{Fig_three_point}  and Fig.~\ref{Fig_four_point} we show examples of loops, cactus and non-cactus  products.
We shall not discuss the averages of non-cactus products here, but let us note that we need to calculate them if we wish 
to determine the whole probability distribution $P(|A_{ij}|)$, through expectations of the moments  $\langle |A_{ij}|^{n} \rangle$.


\vspace{.2cm}

{\bf Correlations}

\vspace{.2cm}

The functions ${\bf F}^{(n)}$  are closely related to  the  time  correlations  \cite{mss}:
\begin{eqnarray}
 O_\beta (t_1,...,t_n) &=& {\mbox{Tr}}\left[  A(t_1) ... e^{-\frac \beta n H}  A(t_n) e^{-\frac \beta n H} \right] / Z \;\nonumber \\
 &=& \left\{\sum_{i_1,...,i_n } e^{-\frac \beta n (e_{i_1}+...+e_{i_n})} A_{i_1 i_2} A_{i_2 i_3} ... A_{i_n,i_1} 
e^{it_1(e_{i_1}-e_{i_2})+it_2(e_{i_2}-e_{i_3})+...it_n(e_{i_n}-e_{i_{1}})}\right\}/Z
\label{otoc}
 \end{eqnarray}
which clearly depend only on time-differences. ({\em From here onwards we set $\hbar=1$}).
Time correlations  are invariant with respect to cyclic permutations of indices, a translation of all times $O_\beta(t_1,...,t_n) = O_\beta(t_1+\tau,...,t_n+\tau)$,  and satisfy the `KMS' condition (the generalization of (\ref{kms2})):
\begin{equation}
O_\beta(t_1,...,t_n) = \left[O_\beta(t_n,...,t_1)\right]^*
\label{KMS1}
\end{equation}
The ordinary $n$-point correlations and responses  can be obtained by analytic continuation. Shifting the various times by a suitable imaginary amount:
\begin{equation}
 O_\beta(t_1-i\beta/n,t_2-2i\beta/n,t_3-3i\beta/n,...) = \left\{\sum_{{i_1,...,i_n }} e^{- \beta  e_{i_1}} {A_{i_1 i_2} A_{i_2 i_3} ... A_{i_n,i_1}} 
e^{i e_{i_1}(t_1-t_n) +...+i e_{i_n}(t_n-t_{n-1})}\right\}/Z
\end{equation}
 Let us consider, for the moment, the partial sum of (\ref{otoc}):
 \begin{eqnarray}
D_\beta (t_1,...,t_n) &=& \left\{\sum_{\stackrel{i_1,...,i_n }{different}} e^{-\frac \beta n (e_{i_n}+...+e_{i_1})} A_{i_1 i_2} A_{i_2 i_3} ... A_{i_n,i_1} 
e^{it_1(e_{i_1}-e_{i_2})+it_2(e_{i_2}-e_{i_3})+...it_n(e_{i_n}-e_{i_{1}})}\right\}/Z 
\label{suma}
\end{eqnarray}
Eq. (\ref{KMS1})  clearly implies also  that $D_\beta(t_1,...,t_n) = \left[D_\beta(t_n,...,t_1)\right]^*$ 
and
 \begin{equation}
 D_\beta(t_1-i\beta/n,t_2-2i\beta/n,t_3-3i\beta/n,...) = \left\{\sum_{\stackrel{i_1,...,i_n }{different}} e^{- \beta  e_{i_1}} {A_{i_1 i_2} A_{i_2 i_3} ... A_{i_n,i_1}} 
e^{i e_{i_1}(t_1-t_n) +...+ie_{i_n}(t_n-t_{n-1})}\right\}/Z \ .
\end{equation}
We also consider  the Fourier transform:
\begin{equation}
\hat D_{\beta} (\omega_1,...,\omega_{n-1}) = 
{ \frac{1}{(2\pi)^{(n-1)}}}
\int dt_1,...,dt_{n-1} \; D_\beta (t_1,...,t_{n-1},0) e^{-i(\omega_1 t_1+...+\omega_{n-1}t_{n-1})}
\label{fourier1}
\end{equation}
which may be written as:
 \begin{eqnarray}
&Z&\hat D_{\beta} (\omega_1,...,\omega_{n-1}) = \nonumber\\ & & \sum_{\stackrel{i_1,...,i_n}{ different}}
e^{-\frac \beta n (e_{i_1}+...+e_{i_n})} A_{i_1 i_2} A_{i_2 i_3} ... A_{i_n,i_1}  \delta(e_{i_1}-e_{i_2}-\omega_1)\delta(e_{i_2}-e_{i_3}-\omega_2)...\delta(e_{i_{n-1}}-e_{i_n}-\omega_{n-1})
\label{EqDomega}
\end{eqnarray}
Eq. (\ref{EqDomega})  describes a function $\hat D_{ \beta}$  made of many deltas of random amplitudes, a `comb'. The definition implies no assumption.
If we now perform an average {of the operators in the spirit of a typicality argument (see next section)} we obtain, assuming (\ref{etheth}):
 \begin{eqnarray}
& & \overline{\hat D_{\beta}} (\omega_1,...,\omega_{n-1}) = \nonumber\\ & & \sum_{\stackrel{i_1,...,i_n}{ different}}
e^{-\frac \beta n (e_{i_1}+...+e_{i_n})} \overline{A_{i_1 i_2} A_{i_2 i_3} ... A_{i_n,i_1}}  \delta(e_{i_1}-e_{i_2}-\omega_1)\delta(e_{i_2}-e_{i_3}-\omega_2)...\delta(e_{i_{n-1}}-e_{i_n}-\omega_{n-1})/Z \label{unouno}\\ & & \sum_{\stackrel{i_1,...,i_n}{ different}} {\cal{N}}^{(n)} (e_+) e^{- \beta e_+} \; {\bf F}^{(n)}_{e_+} (\omega_1,...,\omega_{n-1})
\delta(e_{i_1}-e_{i_2}-\omega_1)\delta(e_{i_2}-e_{i_3}-\omega_2)...\delta(e_{i_{n-1}}-e_{i_n}-\omega_{n-1})/Z \label{dosdos}
\label{suma2}
\end{eqnarray}

Smoothing over a scale of many eigenvalues, but small energy difference $\Delta$,  replacing sums by integrals (see e.g., Ref \cite{srednicki_99,kafri})
$
\sum_i \; \bullet \;\;\; \rightarrow \int de \; e^{S(e)} \; \bullet
$, 
and performing saddle point integration for $e_+$ , we obtain
 \begin{eqnarray}
& &  \overline{\hat D_{\beta} (\omega_1,...,\omega_{n-1})}|_\Delta
= {\bf F}^{(n)}_{e_\beta} (\omega_1,...,\omega_{n-1})
\label{suma3}
\end{eqnarray}
where $e_\beta$ is the canonical energy at temperature $1/\beta$, associated with the partition function $Z$.
This justifies the choice  made for ${\cal{N}}^{(n)}$, that precisely cancels the  levels density.
In going from (\ref{bullet}) to (\ref{suma3}) we have used that:
\begin{eqnarray}
& &S(e_{i_1})+...+S(e_{i_n})= S(e_++e_{i_1}-e_+)+...+S(e_++e_{i_n}-e_+)\nonumber\\
&=& n S(e_+) + \beta [ (e_{i_1}-e_+)+...+(e_{i_n}-e_+)] + S^{''}(e_+)  [ (e_{i_1}-e_+)^2+...+(e_{i_n}-e_+)^2]/2+...\nonumber\\
&\sim&  n S(e_+)
\end{eqnarray}
where we have used that the quadratic and higher terms in $(e_{i}-e_+)$ become negligible -- see argument leading to (210) in \cite{kafri}. Note
 the simplifying role of using a  split  Gibbs-Boltzmann factor, guaranteeing the cancellation of terms linear in  $(e_{i}-e_+)$.

We can also write
\begin{equation}
D_\beta (t_1,...,t_{n-1},0) \sim \int d\omega_1,...,d\omega_{n-1} \;  e^{i(\omega_1 t_1+...+\omega_{n-1} t_{n-1}  )}{\bf F}^{(n)}_{e_\beta} (\omega_1,...,\omega_{n-1})
\label{fourier}
\end{equation}
where the $\sim$ sign means that the approximate  equality holds for  times  $t \ll \Delta^{-1}$, so
that the Fourier transform implicitly smooths the amplitudes of the delta peaks  in Eq.~(\ref{EqDomega}) and we may replace  $\hat D$ by ${\bf F}$ : this allows us to write the equation for $D_\beta$ 
without any other averaging.

In order to reconstruct the complete time correlation (\ref{otoc}), we need to consider all the different possible ways in which indices may coincide. We shall do this in detail
for three and four point functions below, let us state here the basic principle.
Suppose we are calculating:
 \begin{eqnarray}
D_\beta^{(\bullet)} (t_1,...,t_n) &=& \left\{\sum_{\stackrel{i_1,...,i_n }{different}} e^{-\frac \beta n (e_{i_1}+...+e_{i_n})} \underbrace{A_{i_1 i_2} A_{i_2 i_3} ... A_{i_{r-1}i_1}}  \underbrace{A_{i_1i_{r+1}} ... A_{i_n,i_1}} 
e^{it_1(e_{i_1}-e_{i_2})+it_2(e_{i_2}-e_{i_3})+...it_n(e_{i_{n}}-e_{i_1})}\right\}/Z \nonumber \\
&=& \left\{\sum_{\stackrel{i_1,...,i_n }{different}} e^{-\frac \beta n (e_{i_1}+...+e_{i_n})} \underbrace{A_{i_1 i_2} A_{i_2 i_3} ... A_{i_{r-1}i_1}}  \underbrace{A_{i_1i_{r+1}} ... A_{i_n,i_1}} 
e^{i e_{i_1}(t_1-t_n) +...+ie_{i_n}(t_{n}-t_{n-1})}\right\}/Z
\label{sumax}
\end{eqnarray}
as in (\ref{cuatro}). We may replace this by the   product average (\ref{cuatrocuatro}) and compute everything in terms of lower correlation functions. Another
way of doing this is to observe that the if we add in (\ref{otoc}) to the times  $(t_1+s, t_2+s, ..., t_{r-1}+s, t_{r},...,t_{n-1}, t_{n})$, this produces a modification only in
a factor $e^{is(e_{i_{r}}-e_{i_1})}$. Letting $s \rightarrow \infty$ will  produce a rapidly oscillating factor whose contribution will cancel through random phases
-- the usual argument for the diagonal approximation. Hence, we may make the association   $\lim_{s \rightarrow \infty} e^{is(e_{i_{1}}-e_{i_r})} \leftrightarrow \delta_{i_r i_1}$, and:


 \begin{equation}
 \begin{array}{ll}
 & \displaystyle  \lim_{s \rightarrow \infty} O_\beta (t_1+s,t_2+s,...t_{r-1}+s, t_{r},...,t_n) = \, \,\,\,\,\,\, \,\,\,\,\,\, \,\,\,\,\,
\\ \vspace{-0.2cm} \\
& \, \,\,\,\,\, \displaystyle  
\left\{\sum_{i_1,...,i_n } \ \underline{\delta_{i_r i_1} }\ e^{-\frac \beta n (e_{i_1}+...+e_{i_n})} A_{i_1 i_2} A_{i_2 i_3} ... A_{i_n,i_1} 
e^{it_1(e_{i_1}-e_{i_2})+it_2(e_{i_2}-e_{i_3})+...it_n(e_{i_n}-e_{i_{1}})}\right\}/Z
\end{array}
\label{sumaxx}
\end{equation}

As a final remark, let us  mention that if $A$ is an observable satisfying ETH, consistently, also $A^n$ should also be.
Let us here only argue that the $n$-element correlations of $A$ contribute to   $\langle A^n \rangle$:
\begin{eqnarray}
\sum_i (A^n)_{ii} \; e^{-\beta e_i}/Z
&=&O_\beta(-i\beta/n,-2i\beta/n,-3i\beta/n,...) =D_\beta(-i\beta/n,-2i\beta/n,-3i\beta/n,...) + {\mbox{ other}}  \nonumber \\
&\sim&  \int d\omega_1,...,d\omega_{n-1} \;  e^{-i (i\beta \omega_1/n, +2i \beta \omega_2 /n...+i(n-1)\beta\omega_{n-1}/n   )}{\bf F}^{(n)}_{e_\beta} (\omega_1,...,\omega_{n-1})+ {\mbox{ other}}
\end{eqnarray}
so we see that ${\bf F}^{(n)}_{e_\beta}$ definitely has a contribution to the diagonal part of a power of matrices.
The full closure of the forms of expectations under matrix products is an interesting question to check in detail, but we shall do this in a future work.

\section{`Typicality' of operators}\label{Sec_tipicality}

In this section we use `typicality' arguments to justify the claim made above, that
only expectations with every index repeated are non-zero.

\subsection{Two-point functions}

Consider a situation where   $A$ would be a full $M*M$ {\em complex Hermitean} matrix, in the basis of the Hamiltonian (i.e., the function $F$ would be a constant).
A notion of `typicality' that goes back to Von Neumann \cite{neumann}, and further discussed in \cite{goldstein,reimann1} (here we follow \cite{reimann2})
is to assume that $ A $ may be replaced by $A^u= U A U^\dag$, with $U$ a Gaussian random matrix, without altering results such as correlations.  The group associated with $U$
depends on the symmetry of $A$: if it is complex hermitean the $U$ is a unitary matrix, while if $A$ is real it is orthogonal.
We shall here restrict ourselves to the unitary case, the generalization to orthogonal or symplectic cases is straightforward.

This typicality cannot apply for rotations in the entire Hilbert space in reality, because they would destroy the band structure given by $F$ that determines the time-correlation functions, 
but let us ignore this for the moment. We can then evaluate the correlation function as an average over $U$ using:
\begin{equation}
\overline{A^u_{ij} A^u_{ji}}= a \delta_{ij} + b
\label{cf}
\end{equation}
where $a,b$ are related to the invariants of the matrix, $M(a+b)= {\mbox{Tr}} A$ and $M (a+b)^2+M(M-1) b^2 =  {\mbox{Tr}} A^2$.

Clearly, one may generalize this to $n$ point functions \cite{reimann2}:
\begin{equation}
\overline{A^u_{ i_1  j_1} ... A^u_{ i_n  j_n}}= \overline {U_{ i_1 \bar i_1} ... U_{ i_n \bar i_n} U^*_{ j_1 \bar j_1} ... U^*_{ j_n \bar j_n}}
A_{ \bar i_1 \bar j_1} ... A_{\bar i_n \bar j_n}
\end{equation}
where we assumed implicit summation over repeated indices.
For the unitary group \cite{beenaker,reimann2}
\begin{equation}
\overline {U_{ i_1 \bar i_1} ... U_{ i_n \bar i_n} U^*_{ j_1 \bar j_1} ... U^*_{ j_n \bar  j_n}} = \sum_{\sigma, \bar  \sigma} {\cal {C}}(\sigma, \bar  \sigma) \; \delta_{ i_1  \sigma({ j}_1)} ... \delta_{ i_n  \sigma({ j}_n)}
 \delta_{\bar  i_1 \bar  \sigma(\bar j_1)} ... \delta_{ \bar i_n \bar  \sigma(\bar j_n)}
\end{equation}
where $\sigma,  \bar \sigma$ are all permutations and $ {\cal {C}}(\sigma,  \bar \sigma) $ known combinatorial numbers. Therefore:
\begin{equation}
\overline{A^u_{ i_1  j_1} ... A^u_{ i_n  j_n}}=
\sum_{\sigma, \bar  \sigma} {\cal {C}}(\sigma, \bar  \sigma) \; A_{ { \bar \sigma} (\bar j_1) \bar  j_1} ... A_{ { \bar \sigma}(\bar j_n) \bar  j_n} \; \delta_{{  i}_1  \sigma({ j}_1)} ... 
\delta_{{ i}_n  \sigma({ j}_n)} = \sum_{\sigma} {\cal {A}}(\sigma)  \; \delta_{{  i}_1  \sigma({ j}_1)} ... 
\delta_{{ i}_n  \sigma({ j}_n)} 
\end{equation}
The matrix $A$ then enters in the average only through its invariants ${\cal{A}}$, implicitly defined above. {Terms which are not invariant under rotation would be in fact 
be modified by averaging.}
For $n=2$, for instance, there are two terms proportional to $\delta_{i_1 j_1}\delta_{i_2j_2}$ and
$\delta_{i_1j_2} \delta_{i_2 j_1}$ which for the particular case of Eq (\ref{cf}) reduces to
 $\delta_{ i  j}*\delta_{ i  j}=\delta_{ i  j}$ and to $\delta_{ i  i}*\delta_{ j  j}=1$.

In our case, we have to deal with averages of the particular kind:
\begin{equation}
\overline{A^u_{ i_1  i_2} A^u_{ i_2  i_3}  ... A^u_{ i_n  i_1}} = \sum_{\sigma} {\cal {A}}(\sigma)  \; \delta_{{  i}_1  \sigma({ i}_1)} ... 
\delta_{{ i}_n  \sigma({ i}_n)} 
\label{smec}
\end{equation}
where the ${\cal{A}}(\sigma)$ are functions of the invariants of the matrix $A$.
Clearly, the product $\delta_{  i_1  \sigma({ i_1})}... 
\delta_{ i_n  \sigma({ i_n})} $ impose the equalities of  indices in groups. For example $\delta_{i_1 i_2}\delta_{i_2 i_3}...\delta_{i_n i_1}$ imposes the 
equality of all indices, while $\delta_{i_1 i_1}\delta_{i_2 i_2} ... \delta_{i_n i_n}=1$ imposes nothing.

Now, suppose we wish to compute  using the `typicality' of operators a quantity like (\ref{suma}), for the moment still for the case of a `full' $M*M$ random matrix.
We should inject (\ref{smec}) into (\ref{otoc}). 
 \begin{eqnarray}
Z \; O_\beta(t_1,...,t_n) &\rightarrow & \sum_{i_1,...,i_n}\sum_{\sigma} {\cal {A}}(\sigma)  \; \delta_{{  i}_1  \sigma({ i}_1)} ... 
\delta_{{ i}_n  \sigma({ i}_n)} 
e^{it_1(e_{i_1}-e_{i_n})+it_2(e_{i_2}-e_{i_1})+...it_n(e_{i_{n}}-e_{i_{n-1}}) -\frac \beta n (e_{i_1}+...+e_{i_n})}
\label{suma1}
\end{eqnarray}
%
%
 For the case $n=2$,  equation (\ref{suma1}) reads:
 \begin{equation}
 O_\beta(t,t') = {\cal{A}}_1 \sum_{i \neq j} e^{i(t-t') (e_i-e_j)} {e^{- \beta (\epsilon_i+\epsilon_j)/2}} + {\cal{A}}_0 
 \label{typ1}
\end{equation}
where we have explicitly put all diagonal contributions in the second term.
 
\begin{figure}
\centering \includegraphics[angle=0,width=5cm]{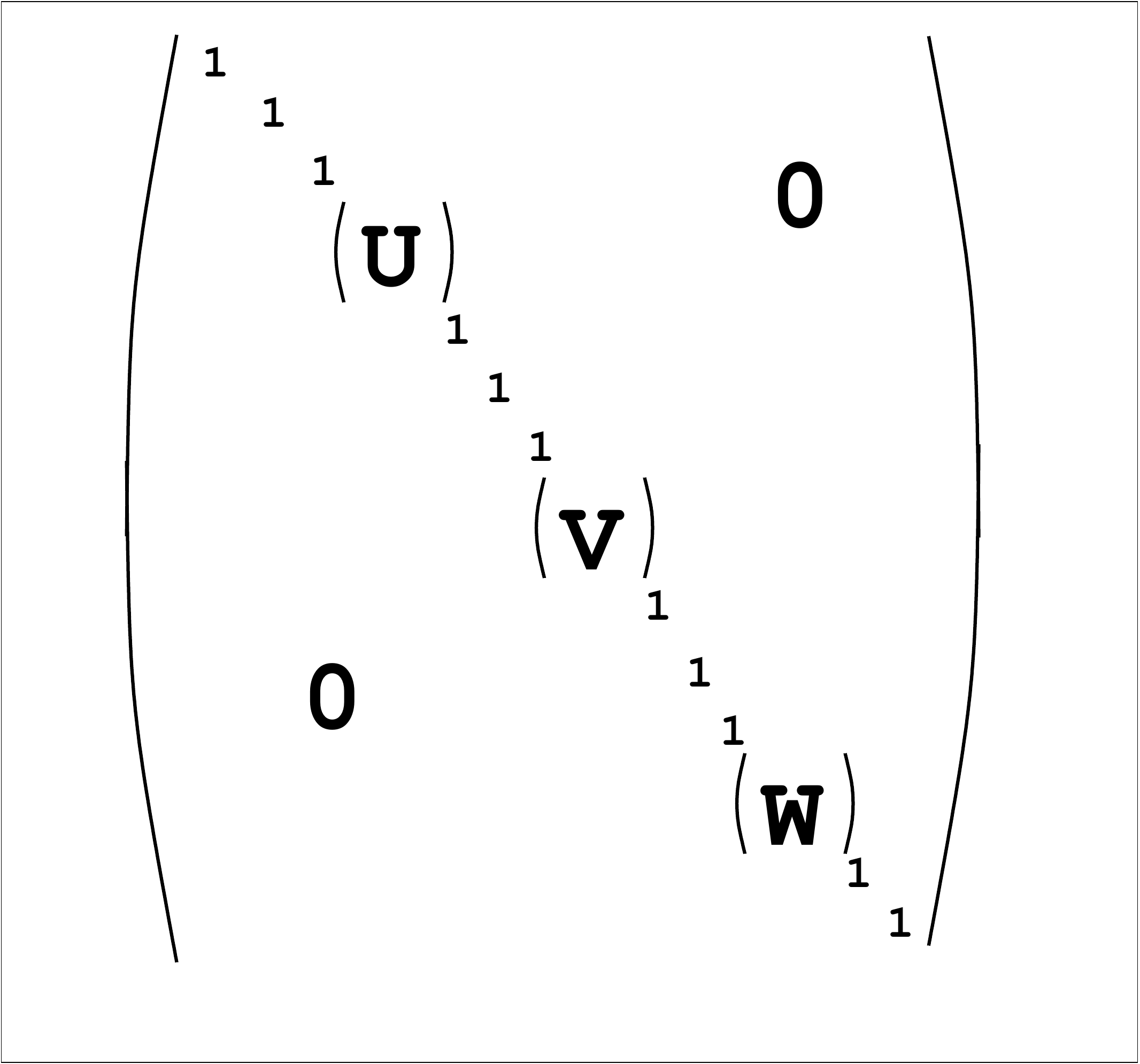}
\caption{A rotation group with three independent $\Delta*\Delta$ unitary transformation matrices}
\label{matrix} 
\end{figure}

 It is clear then, comparing with (\ref{ETH}),  what we are missing in this first `typicality' approach: we need to impose the fact that local operators may be `typical' with respect to rotations 
 between eigenstates associated with nearby levels, but `in the large' they have a band structure given by $F(e_+, \omega)$, which so far is absent. One way to accomplish this \cite{deutsch_91}
has already been mentioned above: it is to add to the Hamiltonian a small ($O(\Delta)$)  random perturbation that will effectively mix level within a range of energy $\Delta$.
We follow here a slightly different route \cite{reimann2}.
Let us then consider  a realistic system, where there is concentration of matrix elements near the diagonal. Imagine for example we wish to estimate which of the
products $A_{ij} A_{kl}$ have a non-zero  expectation value. We first  introduce an energy scale $\Delta$ such that it contains many levels, but is small in the thermodynamic limit
(or, in a semiclassical one).
A modified form of typicality may be introduced as follows: consider the energy intervals
\begin{equation}
\left\{e_i-\frac \Delta 2, e_i+\frac \Delta 2\right\} \;\;\; ; \;\;\; 
\left\{e_j-\frac \Delta 2, e_j+\frac \Delta 2\right\} \;\;\; ; \;\;\;
\left\{e_k-\frac \Delta 2, e_k+\frac \Delta 2\right\} \;\;\; ; \;\;\;
\left\{e_l-\frac \Delta 2, e_l+\frac \Delta 2\right\} 
\end{equation}
and let us first assume that they are disjoint.
We now define a group of unitary matrices of the form Fig. \ref{matrix}, where $U^{(i)},U^{(j)},U^{(k)},U^{(l)}$ are independent $\Delta*\Delta$ matrices.
We now assume that there is typicality within this group, i.e
\begin{equation}
\overline{A_{ij}^u A_{kl}^u} = A_{i'j'}A_{k'l'} \overline{  U^{(around \; i)}_{ii'} \;U^{(around \;j) *}_{jj'}\; U^{(around \;k)}_{kk'}\;U^{(around \;l)*}_{ll'}       }
 \end{equation}
 The average is clearly zero, because each rotation is independent.
 In order to have the possibility of a non-zero average, we need that a matrix $U^{(a)}$ appears at least twice, once conjugated. For this to happen, we need
 that the indices $(i,j,k,l)$ be at most in two intervals. Consider the case when $(i,l)$ belong to an interval, and $(j,k)$ to another, and the rotations associated with these
 two sets are $U$ and $V$ respectively. We have then:
 \begin{equation}
\overline{A_{ij}^u A_{kl}^u} = A_{i'j'}A_{k'l'} \overline{U_{ii'} V^*_{jj'}V_{kk'}U^*_{ll'}}= {\cal{A}}_1({\mbox{intervals}}) \;  \delta_{jk}\delta_{il} 
 \end{equation}
{ Here and in the following  the argument `(intervals)' makes explicit the fact that the constant is dependent on where are located  the intervals considered by the rotations -- i.e. a function of the energies
 defining (up to a small uncertainty  $\Delta$) the two intervals.}
The other possibility is that $(i,j)$ belong to an interval, and $(k,l)$ to another. Then, we have:
 \begin{equation}
\overline{A_{ij}^u A_{kl}^u} = A_{i'j'}A_{k'l'} \overline{U_{ii'} U^*_{jj'}V_{kk'}V^*_{ll'}}= {\cal{A}}_0({\mbox{intervals}}) \;  \delta_{ij}\delta_{kl} 
 \end{equation}
 a product diagonal terms.
 This means that the non-zero expectation is:
 \begin{equation}
\overline{A_{ij}^u A_{ji}^u} = {\cal{A}}_0({\mbox{intervals}}) \;  \delta_{ij} +  {\cal{A}}_1(e_i,e_j) = {\cal{A}}_0({ \epsilon_i}) \;  \delta_{ij} + {\cal{N}}^{(2)} {\bf F}^{(2)}_{e_+}(e_i-e_j) \ .
 \end{equation}
 The normalization ${\cal{N}}^{(2)}$ is for the moment arbitrary. We set it to $ {\cal{N}}^{(2)}= e^{-S(e_+)}$ so that ${\bf F}^{(2)}$ remains of order one. For the first term we assume that $ {\cal{A}}_0=a^2$ may be at most of order one, corresponding to an operator $A$ having a finite
  limit for its large-time two-point function.
 
 \subsection{Three-point functions}
 
Consider first  a term like: $ \overline{ A^u_{i_1,j_1} A^u_{i_2,j_2} ... A^u_{i_n, j_n} } $ or, in particular, $\overline{A^u_{i_1,i_2} A^u_{i_2,i_3} ...  A^u_{i_n,i_1}}$. 
The energies  $e_{i_1} , e_{i_2} ,..., e_{i_n} $  may be contained
in at most $n$ intervals of size $\Delta$.
If we proceed as before, with rotation along at most $n$ intervals around the indices, we find as before that in order to have a non-zero average, a rotation 
 must appear at least twice, one affecting a first and one affecting a second index. This means that   we need to restrict ourselves to $j$ indices such that 
 the $e_j$  are contained in the same intervals.
 In order to see how these averages work, let us develop in detail
 the case $n=3$.

\begin{figure}
\centering \includegraphics[angle=0,width=12cm]{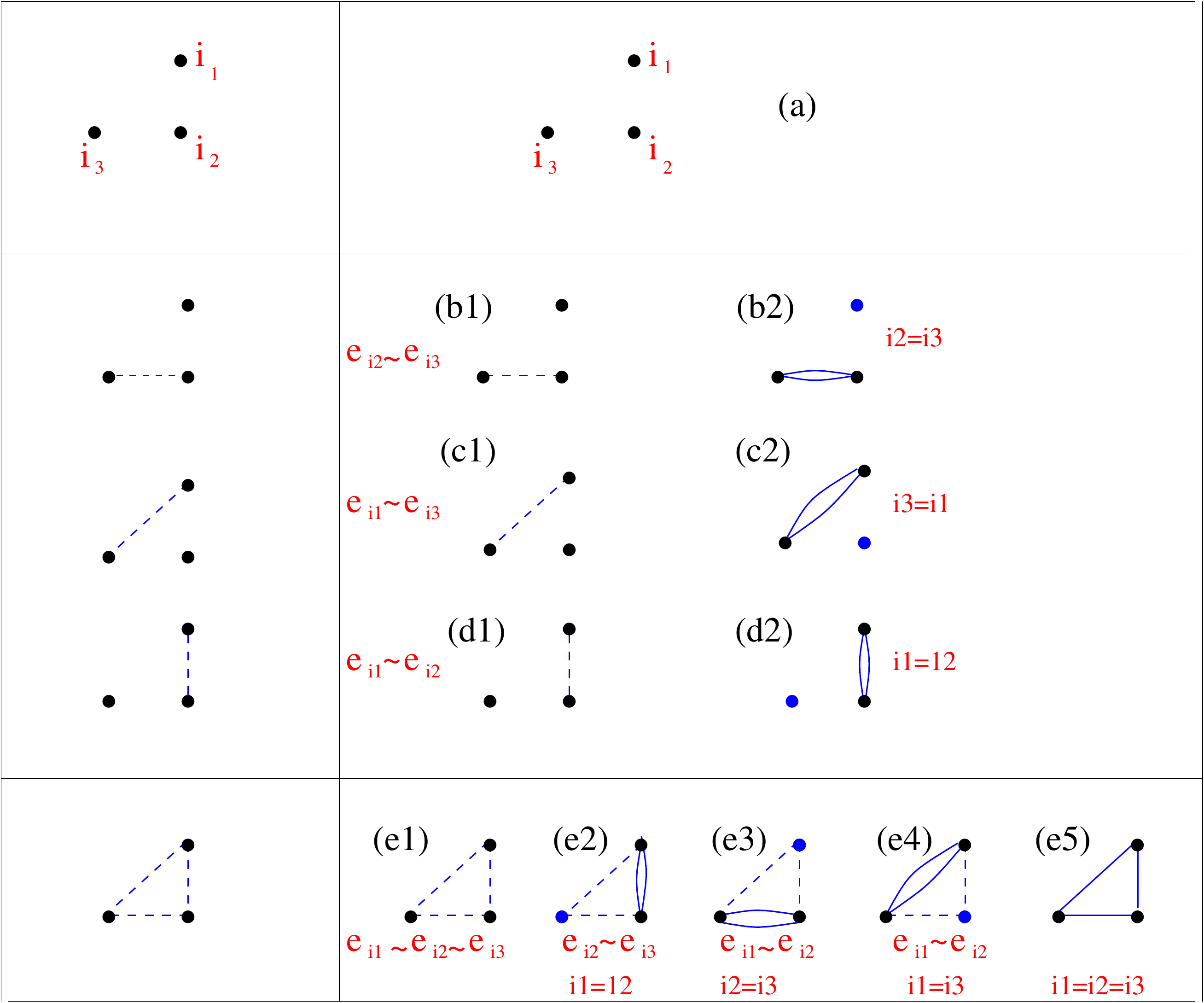}
\caption{A dotted line in the diagram on  the left column means that the indices are in the same energy interval, in a diagram on the right column
that they are in the same interval but are not equal. In the diagrams of the right columns a full line stands for a Kronecker delta, imposing the equality of indices.
A double line imposes the same equality twice (e.g. $\delta_{ij} \delta_{ji}$), while a blued dot a delta with equal indices (e.g. $\delta_{ii}$):  they are indicated only for counting purposes, to show that there are  three Kronecker symbols in each diagram.} \label{diag5} 
 \end{figure}

 Our unitary matrix ensemble is as in Fig. \ref{matrix}. In column to the left of Fig \ref{diag5} we show the combinatorics of the energy intervals. 
 Next, averaging over the independent transformations for every interval, we obtain a number of delta functions imposing some equalities of indices. The diagrams on the right have a full line if indices coincide, a dashed line if they do not coincide but are in the same energy block, and no line when 
 they are in different blocks.
 \begin{itemize}
\item  If two indices are different $k\neq l$, but they belong to the same interval, this means that $|e_k-e_l | \rightarrow 0^+$: they are very close 
in terms of the parameters of the problem $\hbar,  1/N$, but do not coincide.
\item  If two indices are at different intervals, we assume that the average over the combined rotations depends smoothly on the energies $e_k$, $e_l$.
\end{itemize}
 Typicality with respect to random unitary transformation of this kind implies that matrix elements products involving  eigenstates associated to eigenvalues
 that are neighbours or differ by a few level-separations may be completely different from matrix elements products containing the diagonal elements. This is already so
 in the original ETH ansatz, where the diagonal elements $A_{ii}$ scale differently (have a large expectation value) from the off-diagonal elements $A_{ij} \;\; i \neq j$.

Going back to the diagrams \ref{diag5}, the situation is summarised as follows: putting $e_+=(e_i+e_j+e_k)/3$:
\begin{itemize}
\item $(i_1,i_2,i_3)$ in different blocks (diagram (a)) then 
\begin{equation}
\overline{A^u_{i_1,i_2} A^u_{i_2,i_3} A^u_{i_3,i_1} } = {\cal{N}}^{(3)} \;  {\bf F}_{e_+}^{(3)}(e_{i_1}-e_{i_2},e_{i_2}-e_{i_3})
\label{eq3}
\end{equation}
where ${\cal{N}}^{(3)}$ is a smooth function of $e_+$ fixed so that ${\bf F}$ is of $O(1)$. We shall determine it below.
\item   Whenever two energies are different but are in the same block, we assume that the corresponding matrix element is simply given by a limit
of Eq (\ref{eq3}). For example, 
\begin{eqnarray}
&(b1)&:   \hspace{1cm} \overline{A^u_{i_1,i_2} A^u_{i_2,i_3} A^u_{i_3,i_1} } = {\cal{N}}^{(3)} \; \lim_{(e_{i_2}-e_{i_3}) \rightarrow 0^+}{\bf F}^{(3)}_{e_+}(e_{i_1}-e_{i_2},e_{i_2}-e_{i_3})
\nonumber \\
&(e1)&:   \hspace{1cm} \overline{A^u_{i_1,i_2} A^u_{i_2,i_3} A^u_{i_3,i_1} } = {\cal{N}}^{(3)} \; \lim_{(e_{i_1}-e_{i_2}) \rightarrow 0^+}\lim_{(e_{i_2}-e_{i_3}) \rightarrow 0^+}
{\bf F}^{(3)}_{e_+}(e_{i_1}-e_{i_2},e_{i_2}-e_{i_3})
\label{eq3bis}
\end{eqnarray}
This is natural, since the size $\Delta$ of the intervals is many level-separations but otherwise arbitrary.
\item If two indices coincide -- there is a full line in the diagram -- then the elements may scale in a different way, and their form is not obtained as
a limit of ${\bf F}_{e_+}^{(3)}(e_{i_1}-e_{i_2},e_{i_2}-e_{i_3})$.
\end{itemize}

\begin{figure}
\centering \includegraphics[angle=0,width=5cm]{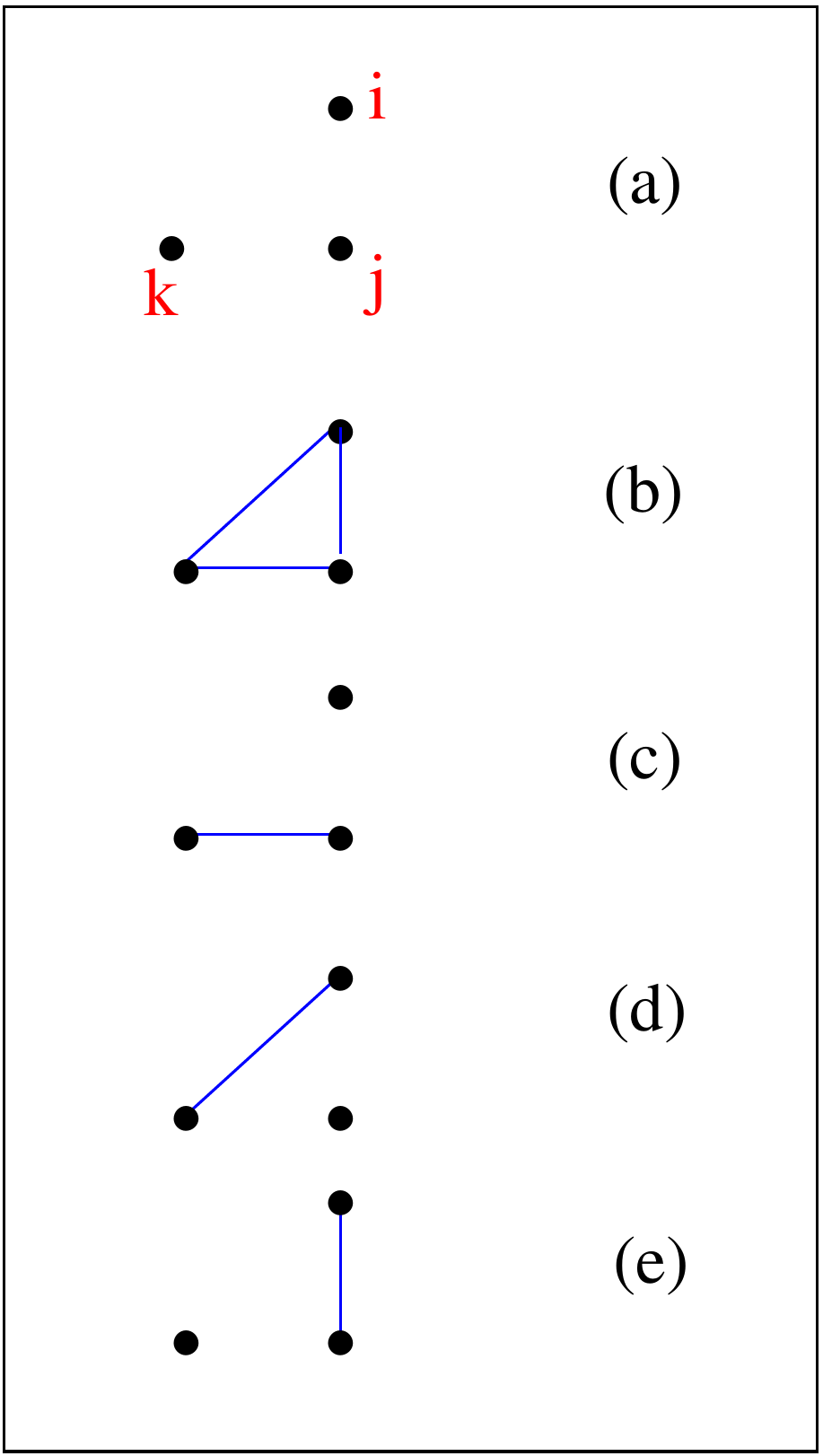}
\caption{The final situation: products as in  diagram (a), diagrams (c,d,e) and diagram (b) have different scaling functions.
}\label{diag3} 
\end{figure}

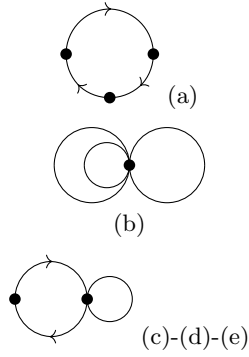
\begin{figure}[h]
\centering

\vspace{0.2cm}
\begin{tikzpicture}[
roundnode/.style={circle, fill=black!100,minimum size=1mm,scale=0.5},
]
\node[roundnode](i0){};
\node[roundnode](i1)[below right=0.66cm of i0] {};
\node[roundnode](i2)[above right=0.66cm of i1] {};
\draw[->] (i0) arc (180:90:0.6cm);
\draw (i2) arc (0:90:0.6cm);
\draw[->] (i2) arc (0:-45:0.6cm);
\draw (i1) arc (270:315:0.6cm);
\draw[->] (i1) arc (270:225:0.6cm);
\draw (i0) arc (180:225:0.6cm);
\end{tikzpicture}
(a)

\vspace{0.2cm}
\begin{tikzpicture}[
roundnode/.style={circle, fill=black!100,minimum size=1mm,scale=0.5},
]
\node[roundnode](i0){};
\draw (i0) arc (360:180:0.5cm);
\draw (i0) arc (0:180:0.5cm);
\draw (i0) arc (180:360:0.5cm);
\draw (i0) arc (180:0:0.5cm);
\draw (i0) arc (360:180:0.3cm);
\draw (i0) arc (0:180:0.3cm);

\end{tikzpicture}

(b)

\vspace{0.2cm}
\begin{tikzpicture}[
roundnode/.style={circle, fill=black!100,minimum size=1mm,scale=0.5},
]
\node[roundnode](i0){};
\node[roundnode](i1)[right=0.8cm of i0] {};
\draw[->] (i0) arc (180:90:0.5cm);
\draw (i1) arc (0:90:0.5cm);
\draw (i0) arc (180:270:0.5cm);
\draw[->] (i1) arc (360:270:0.5cm);
\draw (i1) arc (180:-180:0.3cm);
\end{tikzpicture}
(c)-(d)-(e)

\caption{Products  corresponding to three-point functions  (an alternative representation to Fig.~\ref{diag3}) : a loop and  two cacti.}\label{Fig_three_point}
\end{figure}

All in all, the situation is reduced to Figure \ref{diag3},  with a  scaling function for each set of indices that differ. 
We can now write, for times that are smaller than the Heisenberg time, so that we may replace the matrix products by their averages:
\begin{eqnarray}
O_\beta(t_1,t_2,t_3) &=& {\mbox{Tr}}\left[  A(t_1) e^{-\frac \beta 3 H} A(t_2) e^{-\frac \beta 3 H}  A(t_3) e^{-\frac \beta 3 H} \right] /Z \nonumber\\
&=& \left\{ \overline{\sum}_{i_1,i_2,i_3} e^{-\frac \beta 3 (e_{i_1}+e_{i_2}+e_{i_3})} A_{i_1 i_2} A_{i_2 i_3} A_{i_3,i_1} 
e^{it_1(e_{i_1}-e_{i_3})+it_2(e_{i_2}-e_{i_1})+it_3(e_{i_{3}}-e_{i_2})} \right \} / Z\nonumber\\
&=& \left\{  \sum_{i_1, i_2 , i_3} +  \sum_{i_1,i_1,i_1} +  \sum_{i_1,i_2,i_2} + \sum_{i_1,i_2,i_1}+ \sum_{i_1,i_1,i_3} \right\}/Z \nonumber \\
&\sim& D_\beta(t_1,t_2,t_3) +  D_\beta^{(b)}(t_1,t_2,t_3) +  D_\beta^{(c)}(t_1,t_2,t_3) +  D_\beta^{(d)}(t_1,t_2,t_3) +  D_\beta^{(e)}(t_1,t_2,t_3) 
\end{eqnarray}
In direct correspondence with the diagrams of Fig \ref{diag3} (we have not made explicit the supraindex (a) for the function corresponding to
all indices different: it will be understood by default). 
Here and in what follows, a bar over $\overline{ \sum}$  means that the sum is unrestricted, while a $\sum$ without a bar assumes all indices are
different, unless they have explicitly the same name.

As mentioned above, to clarify the role of the  sum over repeated indices  in the time domain, consider 
the limit of times larger than the Heisenberg time: $\lim_{s \rightarrow \infty} O_\beta (t_1,t_2+s,t_3)$. 
From equation (\ref{otoc}) we see
that the rapidly oscillating factors $e^{is(e_{i_2}-e_{i_3})}$ will cancel everything by random phases, except the diagonal terms with $i_2=i_3$.
Hence 
\begin{equation}
\lim_{s \rightarrow \infty} O_\beta (t_1,t_2+s,t_3)= 
\left\{ {\overline \sum}_{i_1,i_2} e^{-\frac \beta 3 (e_{i_1}+2e_{i_2})} A_{i_1 i_2} A_{i_2 i_2} A_{i_2,i_1} 
e^{i(t_1-t_3)(e_{i_1}-e_{i_2})} \right \} / Z= D_\beta^{(c)} (t_1,t_3) +D_\beta^{(b)} 
\label{clust1}
\end{equation}

\subsection{Four-point functions}

 \begin{figure}
\centering \includegraphics[angle=0,width=7cm]{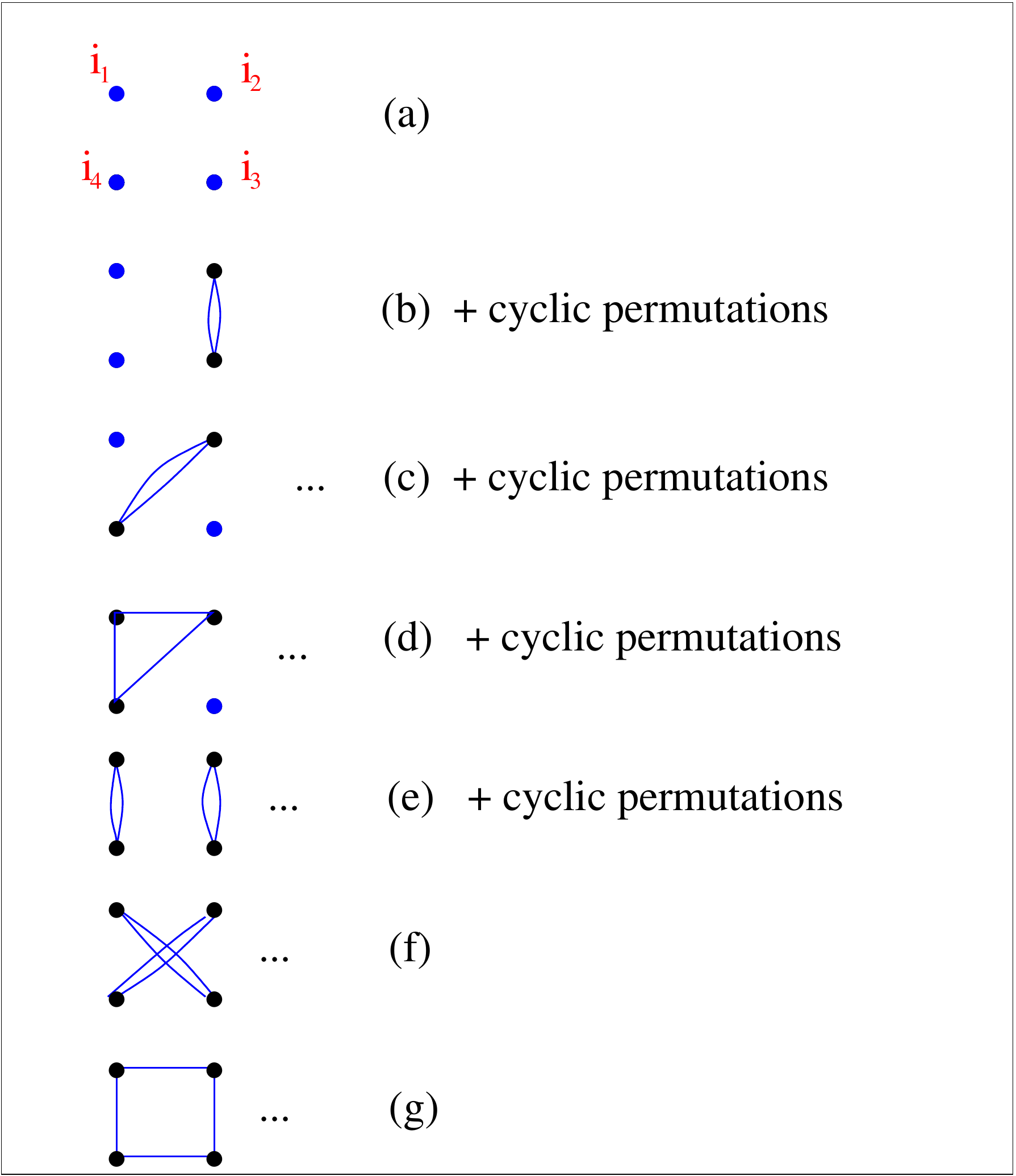}
\caption{Products  corresponding to four-point functions.}\label{diag2} 
\end{figure}

%

We shall consider also the  $4$-point function, as it is the one that is used to define the Lyapunov exponent and the Dynamic Heterogeneity length $\chi_4$
\cite{chi4}:
\begin{eqnarray} Z \; O_\beta(t_1,t_2,t_3,t_4)  &=& {\mbox{tr}} \left[ e^{-\frac{\beta}{4} H}   e^{i t_1  H} A e^{-i t_1  H}        e^{-\frac{\beta}{4} H}   e^{i t_2  H} A e^{-i t_2  H}  e^{-\frac{\beta}{4} H}   e^{i t_3  H} A e^{-i t_3  H}  
e^{-\frac{\beta}{4} H}    e^{i t_4  H} A e^{-i t_4  H}                                                     \right]
\nonumber \\
&=&
\overline{\sum}_{ijkl}e^{-\frac{\beta}{4} e_i}   e^{i (t_1-t_4)  e_i} A_{ij} e^{i (t_2-t_1)  e_j}        e^{-\frac{\beta}{4} e_j}   A_{jk} e^{i (t_3-t_2)  e_k}  e^{-\frac{\beta}{4} e_k}   A_{kl} 
 e^{i (t_4-t_3)  e_l} 
e^{-\frac{\beta}{4} e_l}    A_{li}     
\label{tes}                                     
\end{eqnarray} 
We now wish to express (\ref{tes}) as a sum of terms with all possible index contractions, i.e. corresponding to the diagrams in Fig \ref{diag2}. 
Let us first consider the sum with all indices different, corresponding to diagram (a):
\begin{eqnarray} Z \;  D_\beta(t_1,t_2,t_3,t_4)  &=&{\sum}_{ijkl}e^{-\frac{\beta}{4} e_i}   e^{i (t_1-t_4)  e_i} A_{ij} e^{i (t_2-t_1)  e_j}        e^{-\frac{\beta}{4} e_j}   A_{jk} e^{i (t_3-t_2)  e_k}  e^{-\frac{\beta}{4} e_k}   A_{kl} 
 e^{i (t_4-t_3)  e_l} 
e^{-\frac{\beta}{4} e_l}    A_{li}     
\label{tes1}                                     
\end{eqnarray} 
or its Fourier transform $\hat D_\beta(e_{12},e_{23},e_{34})$ \begin{equation}
 \hat D_\beta(e_{12},e_{23},e_{34})  \equiv \sum_{ijkl} \delta(e_i-e_j-e_{12})\delta(e_j-e_k-e_{23})\delta(e_k-e_l-e_{34})
 e^{-\frac{\beta}{4} e_i}   A_{ij}       e^{-\frac{\beta}{4} e_j}   A_{jk}  e^{-\frac{\beta}{4} e_k}   A_{kl} 
e^{-\frac{\beta}{4} e_l}    A_{li}/Z
\label{ees1}
\end{equation}
Again, the difference between $\hat D_\beta (e_{12},e_{23},e_{34}) $ and $ {\bf F}^{(4)}_\beta(e_{12},e_{23},e_{34})$ is that $\hat D_\beta (e_{12},e_{23},e_{34}) $
is a `comb' of delta-functions, while $ {\bf F}^{(4)}_\beta(e_{12},e_{23},e_{34})$ is a version that has been smoothed either by averaging over an energy window or by
averaging over unitary transformations as described above.
Eq. (\ref{etheth}) is, for the case of four different indices:
\begin{equation}
{\overline{ A_{ij}        A_{jk}     A_{kl}   A_{li}} }=  {\bf F}^{(4)}_{e_+} (e_i-e_j,e_j-e_k,e_k-e_l) e^{-3s(e_+)}
\label{eth4}
\end{equation}

 Other types of contributions to the four point correlation function, corresponding to the diagrams in
 Fig.~\ref{diag2}, are listed in Appendix~\ref{App1}.

\section{The Lyapunov exponent and the out of  time order correlator}

The Lyapunov exponent  is derived from the OTOC  $O_\beta(t,0,t,0)$ by assuming \cite{mss} that there is a time regime, larger than the correlation time, such that:
\begin{equation}
O_\beta(t,0,t,0) \sim C_o - \frac{\hbar^2 C_1}{N^a} e^{\lambda t} + ...  \sim C_o -  e^{\lambda (t-t_e)} 
\label{cococ}
\end{equation}
for times $t_e \gg t \gg 1/\lambda$, where $t_e= \frac{a \ln N/\hbar^2 C_1}\lambda$ is (an estimate of) the Ehrenfest time, the time taken for a minimal packet to spread throughout phase-space \cite{mss,kurchan,ruch} (for a detailed discussion of  the connections between the fidelity decay and the Lyapunov exponent, see Ref. \cite{pasta,PZ}).  The quantum exponent extracted from the OTOC coincides with the classical one in some instances
of the semiclassical limit \cite{kurchan}, but this is not necessarily the case \cite{Galitski}.
It should be also noted in passing, as implicitly stated in Ref. \cite{Galitski}, that the Lyapunov exponent so-defined corresponds to an
average taken {\em before} the logarithm, and not {\em after} as is the normal definition: in the language of disordered systems it
is an "annealed" rather than a "quenched" average.

The terms contributing to the OTOC  can be separated in those resulting from different sums over indices, corresponding
to diagrams of Fig \ref{diag2}. As we have seen above, the only term that is not
directly deducible, with a model-independent~\footnote{This might seem confusing, as in spin-glass models, and in particular SYK model, the four-point function is a functional of the two-point functions. Note, however, that this dependence depends on the model and its parameters, it is not simply a variant of clustering formulas} formula (see below) from two or three-point correlation functions~\footnote{The latter we assume is zero by symmetry.}  is the sum with all indices different, {\em so we expect that the exponential growth 
is given by this term}:
\begin{eqnarray} D_\beta(t,0,t,0)  &=& \frac 1 Z {\sum}_{ijkl}   e^{i (e_i-e_j+e_k-e_l)t} A_{ij}     A_{jk}   A_{kl} A_{li}
e^{-\frac{\beta}{4} (e_i+e_j+e_k+e_l)}         
\label{tes11}                                     
\end{eqnarray} 
a real, even function of time.
In Fourier space it reads:
\begin{eqnarray}
\tilde  G_{\beta}(\omega) &=& \int dt e^{- i\omega t} D_\beta(t,0,t,0)=  \frac 1 Z \sum_{ijkl}  \delta(e_i-e_j+e_k-e_l-\omega)
 e^{-\frac{\beta}{4} e_i}   A_{ij}       e^{-\frac{\beta}{4} e_j}   A_{jk}  e^{-\frac{\beta}{4} e_k}   A_{kl} 
e^{-\frac{\beta}{4} e_l}    A_{li} 
\label{fourier}
\end{eqnarray}
The smoothed version of this reads:
\begin{equation}
G_{\beta} (\omega)  = \overline{\tilde  G_{\beta}(\omega)}|_\Delta= \int  {\bf F}_\beta^{(4)}(\omega-\omega',\omega'', \omega') d\omega' d \omega''
\label{GG}
\end{equation}
 
What to expect of the general features of (\ref{GG})? First of all, because $ D_\beta(t,0,t,0) $ is expected to be nonzero for times smaller than the Ehrenfest time,
this means that $G_{\beta} (\omega)$ should be smooth on a scale $\delta \omega \ll 1/t_e$  (by which we mean that convoluted with a Gaussian 
having a width smaller than that, it remains unaltered). We expect, however,  oscillations of $\delta \omega \sim 1/t_e$, if $D$ extends to $t_e$. The information we seek is in the envelope of these oscillations.

A rapidly oscillating function suggests that one has to study it for complex variables, which leads us 
to compute the Laplace transform:
\begin{eqnarray}
{\cal{G}}(s) &=& \int_0^\infty  dt \; e^{-ts}  D_\beta(t,0,t,0) = \int d\omega \int_0^\infty  dt \; e^{-ts+i\omega t} \tilde  G_{\beta}(\omega) \nonumber\\
&=& \int d\omega \; \frac{s \; \tilde G_{\beta}(\omega)}{s^2+\omega^2}\nonumber\\
&=& \frac s Z \sum_{ijkl} 
 \frac{e^{-\frac{\beta}{4} e_i}   A_{ij}       e^{-\frac{\beta}{4} e_j}   A_{jk}  e^{-\frac{\beta}{4} e_k}   A_{kl} 
e^{-\frac{\beta}{4} e_l}    A_{li} }{ (e_i-e_j+e_k-e_l)^2+s^2}\label{analy} \\
&=& \int d\omega \; \frac{s \;  G_{\beta}(\omega)}{s^2+\omega^2}
\end{eqnarray}
In the last equality we have substituted the function for the smoothed version, a safe thing to do for $s$ much larger than the level spacing.

The function ${\cal{G}}(s)$ is analytic outside the imaginary $s$ axis (cfr Eq (\ref{analy})) for $t_e$ finite. However, in the 
limit of diverging Ehrenfest time,
a form (\ref{cococ}) implies that  ${\cal{G}}(s)$ has a  pole in zero  (rounded off at a scale  $s \sim 1/t_e$) and a pole in 
$s = \lambda$ of residue $\sim e^{-t_e \lambda}$,
rounded off at a scale $(s-\lambda) \sim 1/t_e$. The rounding-off of poles re-establishes analyticity for $t_e$ finite.

{\em The  bound to Lyapunov \cite{mss} means that  $\lim_{t_e \rightarrow \infty} e^{st_e} ({\cal{G}}(s)-C_o/s)$ is analytical for { $s>\frac{2\pi}{\beta}$.}}

\subsection{Numericals results}

Following \cite{kafri}, we compute the correlation functions of a one-dimensional model of hard-core bosons with the Hamiltonian 
\beq
\hat{H}= - J \sum_{j=1}^{L-1} \left( \hat{b}^{\dagger}_j \hat{b}^{}_{j+1} + \text{H.c.} \right) +
V\sum_{j<l} \frac{\hat{n}^{}_j \hat{n}^{}_{l}}{|j-l|^3} +g\sum_{j} x_j^2\, \hat{n}^{}_j.
\label{eq:dhcbmodel}
\eeq
The number of bosons was set to be $L/2$. The three terms in this Hamiltonian describe, from left to right, hopping, dipolar interactions, and a harmonic potential. Here, $x_j$ is the distance of site $j$ from the center of a trap.  For $V\neq0$, this model is non-integrable.

\begin{figure}
\centering \includegraphics[angle=0,width=10cm]{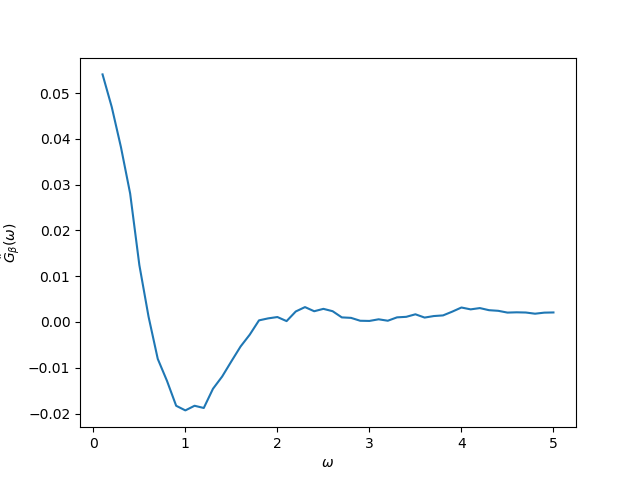}
\caption{The function $\hat G_\beta(\omega)$ for a system of size $N=12$ and where we took $\Delta=0.3$ and $\beta$ corresponding to $\epsilon_+$ in
the middle of the spectrum}
\label{Gw} 
\end{figure}

\begin{figure}
\centering \includegraphics[angle=0,width=10cm]{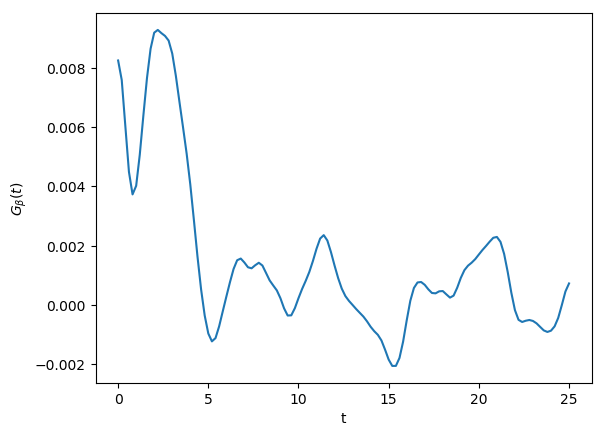}
\caption{The function $G_{\beta}(t)$ for a system of size $N=10$ and where we took $\Delta=0.3$ and $\beta=0$}
\label{Dt} 
\end{figure}

\begin{figure}
\centering \includegraphics[angle=0,width=10cm]{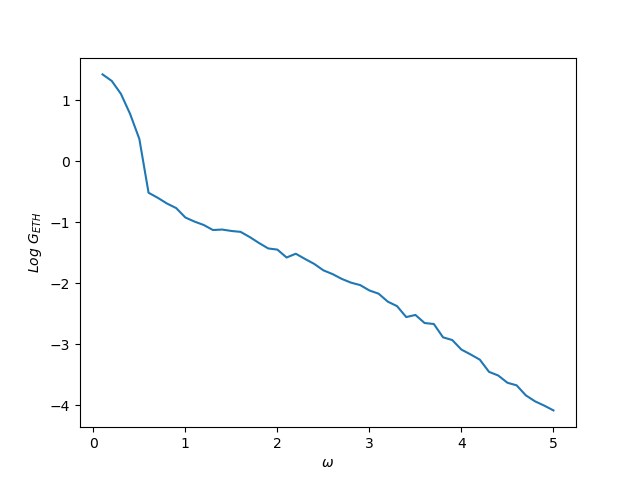}
\caption{The contribution to the OTOC of all terms except the one with all the indices different, the Fourier transform of $O_\beta(t,0,t,0)-D_\beta(t,0,t,0)$. The curve is very similar to the two-point curve,
and is hence unimportant for the Lyapunov  times. 
The system of size $N=12$ and we took $\Delta=0.3$ and $\epsilon_+$ in
the middle of the spectrum.}
\label{GETH} 
\end{figure}

Figure \ref{Gw} shows the function $\hat G_\beta(\omega)$ computed for a system of size $N=12$ where we
considered as an observable the site occupation at the center of the chain, and Fig \ref{Dt} shows its Fourier transform. 
The sizes are small, but we clearly see correlations that are absent in the usual assumption of independence of matrix elements
unrelated by symmetry.
Figure \ref{GETH} shows the contribution of the all terms except the ones in Fig. \ref{Gw}: we see a behaviour very close to that
of the two-point function computed in  Ref \cite{kafri}.

\section{Conclusions}

We have argued that that matrix elements of local operators in the basis of a chaotic Hamiltonian retain correlations that are
small but contribute to higher correlation functions. 
The set of nonzero expectations of products of matrix elements is in a one-to-one relation with the set of 
connected correlator.
Even if the  ETH ansatz with independent elements  may be shown to close under products \cite{srednicki_99,kafri},
so that, for example, the square of an operator of the ETH form is also of the ETH form, the neglected correlations between elements contribute and disregarding them may lead
to  incorrect results. A simple argument to convince oneself of this is that, if expectations of diagonal  and variances of off-diagonal elements sufficed to determine
all $n$-time correlations, then all could be written as  model-independent functional of one and two-time correlations, which is for general models 
obviously not the case.  An interesting generalization of the present work is to derive ETH forms for the joint distribution of various operators, under the same assumptions of this paper.
 
 \section{Acknowledgments}

We acknowledge M. Rigol and Y. Kafri for useful discussions.
L.F. and J.K. are supported by the Simons  Foundation Grant No 454943.

 \appendix
 
 \section{Contributions from other diagrams}\label{App1}

The other contributions to $O_\beta(t,0,t,0)$ associated to the diagrams in Fig.~\ref{diag2}
derive from the  2-point function within the standard ETH or
its extension for the three point function:

\beq
\begin{array}{ll}
\displaystyle
D^{(b) }(t) = \frac{1}{Z} \sum_{i,j,k} e^{-\beta (2 \epsilon_j+\epsilon_k+\epsilon_i)/4} e^{i (\epsilon_i-\epsilon_j)t} e^{i (\epsilon_j-\epsilon_k)t} 
A_{ij} A_{jj} A_{jk} A_{ki}
\\ \vspace{-0.2cm} \\
\displaystyle
\sim \int {\rm d} \epsilon_1 {\rm d} \epsilon_2   e^{\beta(\epsilon_1+\epsilon_2)/12} e^{i( \epsilon_1 - \epsilon_2) t} {\bf F}^{(3)}_{\epsilon_\beta^{} }(\epsilon_1,\epsilon_2) {\bf F}^{(1)}(\epsilon_\beta+3\epsilon_1/4-\epsilon_2/4)
\end{array}
\eeq
where we made the change of variables $\epsilon_1= \epsilon_i-\epsilon_j$, $\epsilon_2=\epsilon_k-\epsilon_j$

\beq
\begin{array}{ll}
\displaystyle
D^{(c) }(t) = \frac{1}{Z} \sum_{i,j,k} e^{-\beta (2 \epsilon_i+\epsilon_k+\epsilon_j)/4} e^{i (\epsilon_i-\epsilon_j)t} e^{i (\epsilon_i-\epsilon_k)t} 
A_{ij} A_{ji} A_{ik} A_{ki}

\displaystyle
 \sim \left[ \int {\rm d} \epsilon_1 e^{-\beta\epsilon_1/4} e^{i \epsilon_1 t} {\bf F}^{(2)}_{\epsilon_\beta}(\epsilon_1) \right]^2
\end{array}
\eeq
with  $\epsilon_1= \epsilon_i-\epsilon_j,\epsilon_i-\epsilon_k$

\beq
\begin{array}{ll}
\displaystyle
D^{(d) }(t) = \frac{1}{Z} \sum_{i,j} e^{-\beta (3 \epsilon_i+\epsilon_j)/4} e^{i  (\epsilon_i-\epsilon_j)t} 
A_{ii} A_{ii} A_{ij} A_{ji}

\displaystyle
\sim \int {\rm d} \epsilon_1 e^{-\beta\epsilon_1/2}  e^{i  \epsilon_1 t} {\bf F}^{(2)}_{\epsilon_\beta}(\epsilon_1) [ {\bf F}^{(1)}(\epsilon_\beta + \epsilon_1/4) ]^2
\end{array}
\eeq
with  $\epsilon_1= \epsilon_i-\epsilon_j$

\beq
\begin{array}{ll}
\displaystyle
D^{(e) }(t) = \frac{1}{Z} \sum_{i,j} e^{-\beta ( \epsilon_i+\epsilon_j)/2} 
A_{ij} A_{jj} A_{ji} A_{ii}
\displaystyle
\sim \int {\rm d} \epsilon_1  {\bf F}^{(2)}_{\epsilon_\beta}(\epsilon_1) {\bf F}^{(1)}(\epsilon_\beta +\epsilon_1/2) 
{\bf F}^{(1)}(\epsilon_\beta - \epsilon_1/2) 
\end{array}
\eeq
with  $\epsilon_1= \epsilon_i-\epsilon_j$

The diagram $(f)$ corresponds to a diagram that is not a cactus and therefore does not contribute (see comment in Sec.~\ref{Sec_matrix_elements}).
Finally one has:

\beq
\begin{array}{ll}
\displaystyle
D^{(g) }(t) = \frac{1}{Z} \sum_{i} e^{-\beta \epsilon_i} 
A_{ii}^4 
%
%
\displaystyle
\sim [ {\bf F}^{(1)}({{\epsilon_\beta} }) ]^4
\end{array}
\eeq

\section{Counting argument}\label{Appendix_entropy}

Let us consider a loop with repetitions characterized by $E=\epsilon_1+\epsilon_2+\dots+\epsilon_n$. 
This is partitioned in $k$ subloops of sizes $n_a$ and energies $E_a=\epsilon_{i_1}+\dots+\epsilon_{i_{n_a}}$ with $a=1,\dots,k$.
We define $\lambda_a=n_a/n$, with $0\leq\lambda_a\leq 1$, $\sum_{a=1}^k \lambda_a=1$, and $x_a=E_{a}/n_a$.
It holds:
\begin{equation}
(n-1) S(\epsilon_+) = (n-1) S(\sum_{a=1}^k \lambda_a x_a)    \geq (n-1)\sum_{a=1}^k \lambda_a S(x_a) \geq \sum_{a=1}^k (n_a-1) S(x_a)
\end{equation}
where in the first inequality we have used the convexity of the entropy and in the second the fact that $\lambda_a\leq1$.
Therefore, if we assume that the covariances of such loops (and their generalization to multi-point functions) scale as $e^{-(n-1) S(\epsilon_+)}$
we see that the expectation of the maximal contraction in loops dominates the expectation of the diagram.


\begin{thebibliography}{100}
\bibitem{foot} Note that the dynamics are not frozen after the Heisenberg time, instead the fidelty can undergo a dephasing dynamics beyond that time,
see Ref. \cite{PZ}.
\end{thebibliography}
\end{document}